\documentclass{ws-procs9x6}

\begin{document}

\title{The Role of Higher Twist in Determining\\
Polarized Parton Densities from DIS data\footnote{\uppercase{T}his
work is supported by the \uppercase{JINR-B}ulgaria
\uppercase{C}ollaborative \uppercase{G}rant, by the \uppercase{
RFBR} (\uppercase{N}o 02-01-00601, 03-02-16816), \uppercase{
INTAS} 2000 (\uppercase{N}o 587) and by the \uppercase{B}ulgarian
\uppercase{N}ational \uppercase{S}cience \uppercase{F}oundation
under \uppercase{C}ontract \uppercase{P}h-1010. }}

\author{E. Leader}

\address{Imperial College London, London WC1E 7HX, England }

\author{A.V. Sidorov}

\address{Bogoliubov Theoretical Laboratory,\\
Joint Institute for Nuclear Research, 141980 Dubna, Russia }

\author{D.B. Stamenov}

\address{Institute for Nuclear Research and Nuclear Energy,
1784 Sofia, Bulgaria \\
E-mail: stamenov@inrne.bas.bg}

\maketitle \abstracts{Different methods to extract the polarized
parton densities from the world polarized DIS data are considered.
The higher twist corrections $h^N(x)/Q^2$ to the spin dependent
proton and neutron $g_1$ structure functions are found to be
non-negligible and important in the QCD analysis of the present
experimental data. Their role in determining the polarized parton
densities in the framework of the different approaches is
discussed. }

One of the features of the polarized DIS is that a lot of the
present data are at low $Q^2$ ($Q^2 \sim 1-5~GeV^2$). For that
reason, to confront correctly the QCD predictions to the
experimental data and to determine the polarized parton densities
a special attention should be paid to the non-perturbative higher
twist (powers in $1/Q^2$) corrections to the nucleon structure
functions. The size of higher twist corrections (HT) to the spin
structure function $g_1$ and their role in determining the
polarized parton densities in the nucleon using different
approaches of QCD fits to the data are discussed in this talk.

Up to now, two approaches have been mainly used to extract the
polarized parton densities (PPD) from the world polarized DIS
data. According to the first \cite{GRSV,LSS2001} the leading twist
LO/NLO QCD expressions for the structure functions $g_1^N$ and
$F_1^N$ have been used in order to confront the data on spin
asymmetry $A_1(\approx g_1/F_1)$ and $g_1/F_1$. We have shown
\cite{LomConf,newHTA1} that in this case the extracted from the
world data ``effective'' HT corrections $h^{g_1/F_1}(x)$ to the
ratio $g_1/F_1$
\begin{equation}
{g_1(x,Q^2)\over F_1(x,Q^2)}={g_1(x,Q^2)_{\rm LT}\over
F_1(x,Q^2)_{\rm LT}} + {h^{g_1/F_1}(x)\over Q^2} \label{A1HT}
\end{equation}
are negligible and consistent with zero within the errors, {\it
i.e.} $h^{g_1/F_1}(x) \approx 0$. (Note that in QCD: $~g_1 =
(g_1)_{LT} + (g_1)_{HT};~F_1 = (F_1)_{LT} + (F_1)_{HT}.~$) What
follows from this result is that the higher twist corrections to
$g_1$ and $F_1$ compensate each other in the ratio $g_1/F_1$ and
the PPD extracted this way are less sensitive to higher twist
effects.

According to the second approach \cite{SMC,BB}, $g_1/F_1$ and
$A_1$ data have been fitted using phenomenological
parametrizations of the experimental data for the unpolarized
structure function $F_2(x,Q^2)$ and the ratio $R(x,Q^2)$ of $F_2$
and $F_1$ ($F_1$ has been replaced by the usually extracted from
unpolarized DIS experiments $F_2$ and $R$). Note that such a
procedure is equivalent to a fit to $(g_1)_{exp}$, but it is more
precise than the fit to the $g_1$ data themselves actually
presented by the experimental groups because the $g_1$ data are
extracted in the same way for all of the data sets.

If the second approach is applied to the data, the ``effective
higher twist'' contribution $h^{g_1/F_1}(x)/Q^2$ to
$A_1(g_1/F_1)$ is found \cite{GRSV} to be sizeable and important
in the fit [the HT corrections to $g_1$ cannot be compensated
because the HT corrections to $F_1(F_2$ and $R)$ are absorbed by
the phenomenological parametrizations of the data on $F_2$ and
$R$]. Therefore, to extract correctly the polarized parton
densities from the $g_1$ data, the HT corrections to $g_1$ have
to be taken into account. Note that a QCD fit to the data in this
case, keeping in $g_1(x,Q^2)_{QCD}$ only the leading-twist
expression (as it was done in \cite{SMC,BB}), leads to some
"effective" parton densities which involve in themselves the HT
effects and therefore, are not quite correct.

Keeping in mind the discussion above we have analyzed the world
data on inclusive polarized DIS \cite{world} taking into account
the higher twist corrections to the nucleon structure function
$g_1^N(x, Q^2)$. In our fit to the data we have used the following
expressions for $g_1/F_1$ and $A_1$:
\begin{eqnarray}
\nonumber \left[{g_1^N(x,Q^2)\over F_1^N(x,
Q^2)}\right]_{exp}~&\Leftrightarrow&~ {{g_1^N(x,Q^2)_{\rm
LT}+h^N(x)/Q^2}\over F_2^N(x,Q^2)_{exp}}2x{[1+
R(x,Q^2)_{exp}]\over (1+\gamma^2)}~,\\
A_1^N(x,Q^2)_{exp}~&\Leftrightarrow&~{{g_1^N(x,Q^2)_{\rm LT}+
h^N(x)/Q^2}\over F_2^N(x,Q^2)_{exp}}2x[1+R(x,Q^2)_{exp}]~,
\label{g1F2Rht}
\end{eqnarray}
where $g_1^N(x,Q^2)_{\rm LT}$ is given by the leading twist QCD
expression including the target mass corrections (N=p, n, d). The
dynamical HT corrections $h^N(x)$ in (\ref{g1F2Rht}) are included
and extracted in a {\it model independent way}. In our analysis
their $Q^2$ dependence is neglected. It is small and the accuracy
of the present data does not allow to determine it. The details
of our analysis are given in \cite{LSSHT}. Unlike the paper
\cite{LSSHT}, the polarized PD determined in this analysis are
compatible with the positivity bounds imposed by the MRST'02
unpolarized PD instead of the MRST'99 ones. The dependence of the
polarized PD on the positivity constraints imposed will be
discussed in detail in a forthcoming paper.

We have found that the fit to the data is significantly improved
when the higher twist corrections to $g_1$ are included in the
analysis, especially in the LO QCD case. The best LO and NLO($\rm
\overline{MS}$ scheme) fits correspond to $\chi^2_{\rm
DF,LO}=0.91$ and to $\chi^2_{\rm DF,NLO}=0.89$, while in the case
of LO and NLO($\rm \overline{MS}$) fits when HT are not included,
$\chi^2_{\rm DF,LO}=1.41$ and $\chi^2_{\rm DF,NLO}=1.19$,
respectively.

\begin{figure}[ht]
\centerline{ \epsfxsize=2.1in\epsfbox{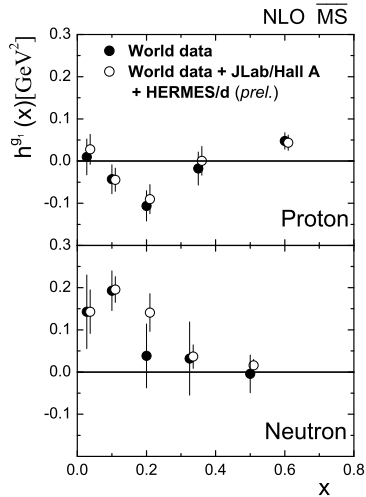}
}
 \caption{
Higher twist corrections to the proton and neutron $g_1$ structure
functions extracted from the data on $g_1$ in the NLO QCD
approximation for $g_1(x,Q^2)_{\rm LT}$. \label{inter}}
\end{figure}

We have also found that the size of the HT corrections to $g_1$ is
{\it not} negligible and their shape depends on the target (see
Fig. 1). In Fig. 1 our results on the HT corrections to $g_1$
(open circles) including in the world data set the recent
JLab/Hall A \cite{JLab} and preliminary HERMES \cite{HERMESd} data
are also presented. As seen from Fig. 1, the higher twist
corrections to the neutron spin structure functions in the large
$x$ region are much better determined now. It was also shown that
the NLO QCD polarized PD($g_1^{\rm LT}+\rm HT$) determined from
the data on $g_1$, including higher twist effects, are in good
agreement with the polarized PD($g_1^{\rm NLO}/F_1^{\rm NLO}$)
found from our analysis of the data on $g_1/F_1$ and $A_1$ using
for the structure functions $g_1$ and $F_1$ only their {\it
leading} twist expressions in NLO QCD. This observation confirms
once more that the higher twist corrections $h^{g_1/F_1}(x)$ to
$g_1/F_1$ and $A_1$ are negligible, so that in the analysis of
$g_1/F_1$ and $A_1$ data it is enough to account only for the
leading twist of the structure functions $g_1$
and $F_1$.\\

In conclusion, we have found that in order to confront the QCD
predictions for the nucleon spin structure function $g_1$ to the
present experimental data on $g_1$ and to extract correctly the
polarized parton densities from these data, the higher twist
corrections to $g_1$ have to be taken into account in the
analysis. While, in the fit to $g_1/F_1$ and $A_1(\approx
g_1/F_1)$ data it is enough to account only for the leading twist
contributions to the structure functions $g_1$ and $F_1$ because
the higher twists corrections to $g_1$ and $F_1$ compensate each
other in the ratio $g_1/F_1$. Further investigations on the role
of higher twist effects in semi-inclusive DIS processes would be
important for the correct determination and flavor separation of
the valence and light sea quark parton densities.

\end{document}